\documentclass[letterpaper, 10 pt, conference]{ieeeconf}  % Comment this line out if you need a4paper

\IEEEoverridecommandlockouts                

\usepackage{graphicx,dblfloatfix} % for pdf, bitmapped graphics files
\usepackage{epsfig} % for postscript graphics files
\usepackage{mathptmx} % assumes new font selection scheme installed
\usepackage{times} % assumes new font selection scheme installed
\usepackage{amsmath} % assumes amsmath package installed
\usepackage{amssymb}  % assumes amsmath package installed
\usepackage{cite}
 \usepackage{tikz}
 \usepackage{graphicx,color,psfrag}
\usetikzlibrary{shapes,arrows,shadows,calc,patterns,decorations.pathmorphing,decorations.markings}
\usepackage{verbatim}
\definecolor{tublue}{RGB}{0,166,214}
\definecolor{tuyellow}{RGB}{247,235,144}
\definecolor{blue}{RGB}{0,166,214}
 \definecolor{lblue}{RGB}{119,192,215}
 \definecolor{lred}{RGB}{236,127,44}
 \newtheorem{proposition}{Proposition}
 \newtheorem{remark}{Remark}
 \newtheorem{assumption}{Assumption}
 
 \newtheorem{problem}{Problem}
\title{A Switching Thrust Tracking Controller for Load Constrained Wind Turbines}

\author{Jean Gonzalez Silva, Daan van der Hoek, Sebastiaan Paul Mulders, Riccardo Ferrari and \\ Jan-Willem van Wingerden$^{*}$ 
\thanks{$^{*}$ Delft University of Technology, Delft, 2628CD The Netherlands {\tt\small \{J.GonzalezSilva, D.C.VanderHoek, S.P.Mulders, R.Ferrari, J.W.vanWingerden\}@tudelft.nl}}%
}

\begin{document}

\maketitle
\thispagestyle{empty}
\pagestyle{empty}

%%%%%%%%%%%%%%%%%%%%%%%%%%%%%%%%%%%%%%%%%%%%%%%%%%%%%%%%%%%%%%%%%%%%%%%%%%%%%%%%
\begin{abstract}

Wind turbines are prone to structural degradation, particularly in offshore locations. Based on the structural health condition of the tower, power de-rating strategies can be used to reduce structural loads at the cost of power losses.
This paper introduces a novel closed-loop switching control architecture to constrain the thrust in individual turbines.
By taking inspiration from developments in the field of reference governors, an existing demanded power tracking controller is extended by a thrust tracking controller. The latter is activated only when a user-defined constraint on fore-aft thrust force  is exceeded, which can be set based on the actual damage status of the turbine. Having a down-regulation with monotonic aerodynamic load response, a simple linear thrust tracking controller is proposed. Such a scheme can reduce aerodynamic loads while incurring acceptable losses on power production which, in a wind farm setting, can be compensated for by other turbines.
Large eddy simulations demonstrate the performance of the proposed scheme on satisfying thrust constraints.
 
\end{abstract}

%%%%%%%%%%%%%%%%%%%%%%%%%%%%%%%%%%%%%%%%%%%%%%%%%%%%%%%%%%%%%%%%%%%%%%%%%%%%%%%%
\section{INTRODUCTION}

%\emph{- Problem definition}

To make wind energy competitive in the transition from fossil fuel-based to renewable energy sources, it is essential to reduce the Levelized Cost of Energy (LCoE). This performance indicator takes into account the costs of construction, maintenance and the energy generated by the power plant over its entire lifetime \cite{ashuri2014}. 

Structural degradation reduces the lifetime of turbine components as an inevitable result of alternating stresses and environmental conditions, thereby increasing the cost of maintenance. Periodic structural loading is  being considered as the main cause for component failures \cite{sutherland1999, bossanyi2003}. Additionally, structural degradation is accelerated by the offshore environment, including wave action and corrosion\cite{price2017}. 

%\emph{- Background \ Motivation}
A way to reduce the damage propagation and increase reliability is by derating the turbine appropriately to reduce structural stresses. Health monitoring systems can be used to detect and to estimate high level corrosion, mechanical flaws and cracks~\cite{may2015,liu2021a}. Although this results in sub-optimal power generation for individual turbines, the turbines' structural reliability is improved, where fatigue damage is alleviated and lifetime extended \cite{griffith2014}.
As a result, the turbine is able to continue operating as opposed to shutting down, until maintenance is fully performed. 

When the goal of the wind farm controller is to track a power reference, and enough power is available in the wind, the power contributions from individual turbines can be redistributed among the turbines in the farm. That is,
reducing the demanded power for a set of turbines, is compensated for by increasing the power reference for the other turbines. This can be accomplished by simple compensation control loops as described in \cite{vanWingerden2017}. For instance, power losses due to waked conditions are compensated when turbines are not able to meet their individual power reference in \cite{silva2021}. 

At an individual turbine level, the down-regulation of power can be achieved by either operating the turbine at higher rotor speeds (HRS) or lower rotor speeds (LRS),
with respect to the rotor speed from conventional controllers.

HRS approaches are beneficial for power tracking because there is more kinetic energy associated with the higher rotor speed, allowing fast recovery when demanded power increases. As an HRS approach, the Max-$\Omega$ strategy \cite{mirzaei2014} maximizes the rotor speed. %Equivalently, the combination of constant blade pitch and constant tip-speed ratio, $\lambda$, produces similar results \cite{kim2018}. 
%However, associated with the HRS approaches, the 1P, 3P and higher harmonic frequencies are increased. Although the HRS approaches alleviate the drive-train loads, they may increase fatigue loads on the support structure by operating near one of the tower modes and by increasing the amount of load cycles. Also,
However, the HRS approaches present limitations for our design. The aerodynamic loads do not reduce significantly when the power demand decreases and might even increase \cite{vanderhoek2018}. As a result of the latter, the thrust response might not be monotone to power setpoint changes. Non-monotone behaviors do not only appear in the thrust and power relation, but they can also appear in the demanded power and generated power relation \cite{deshpande2012}.

\begin{figure*}[b!]
\centering
\includegraphics[width=0.8\linewidth]{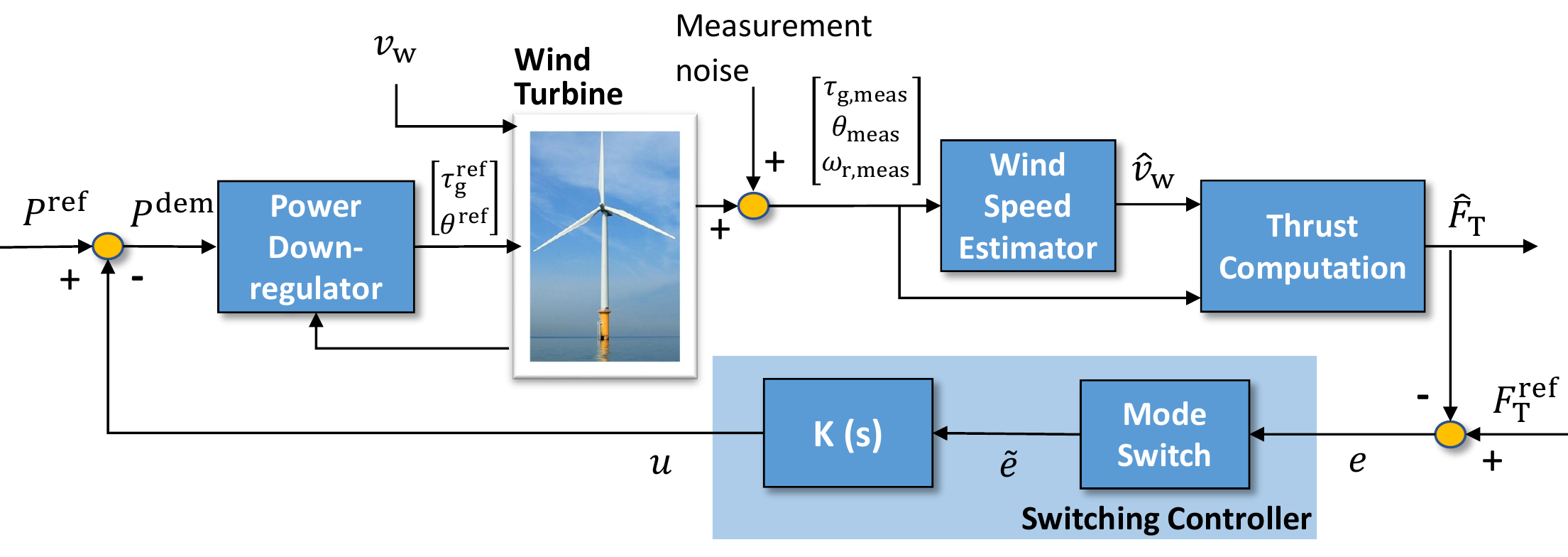}
\caption{Block diagram of the proposed switching control architecture for constraining the turbine thrust forces. The novel switching controller is highlighted in light blue.}
\label{clthrust}
\end{figure*}

LRS approaches present lower aerodynamic loads and, consequently, lower structural loads on the blades and tower. In constrast with HRS approaches, the turbine response presents a monotone behavior from power demand to thrust. As an LRS approach, the Const-$\Omega$ \cite{lio2018} keeps the rotor speed constant avoiding high rotor speeds, but can reach extreme tip-speed ratio values and undesirable operation regions as wind speed changes. Furthermore, there exists the so-called min-C$_\mathrm{T}$ method based on steady state power coefficient (C$_\mathrm{P}$) and thrust coefficient (C$_\mathrm{T}$) look-up tables \cite{lio2018,juangarcia2018}. C$_\mathrm{P}$ and C$_\mathrm{T}$ represent the wind turbine power and thrust conversion efficiencies, respectively. %Then, a unique pair of desirable tip-speed ratio and blade pitch, [$\lambda_\mathrm{d}$, $\theta_\mathrm{d}$], can be derived from the min-C$_\mathrm{T}$/max-C$_\mathrm{P}$ curve. 
The method ideally leads to the smallest thrust compared to all the down-regulation strategies and demonstrates benefits for waked wind farms, which allows more available power to down-stream turbines \cite{santoni2015,ma2017}. %The demanded power $P^\mathrm{dem}$ is tracked by the corresponding power conversion efficiency  
%$\text{C}_\mathrm{P, \, d}= P^\mathrm{dem}/\eta_\mathrm{eff}0.5 \rho \pi R^2 v_\mathrm{w}^3,$
%where $\eta_\mathrm{eff}$ is the total efficiency of the turbine, $\rho$ the air density, $R$ the rotor radius and $v_\mathrm{w}$ the wind speed.  
However, its operation in low tip-speed ratios is close to stall conditions, where the predominant flow stability along the blades can be lost, affecting the controlled turbine response.
This latter drawback is overcome in \cite{aho2016} and \cite{kim2018} with the so-called ``Active Power Control (APC) pitch" and KNU2, respectively. This approach pitches the blades to follow a reference generator speed, based on the function built from the conventional generator torque law \cite{jonkman2009,Mulders2020a}. Although the power tracking response is slightly worse in the LRS approaches \cite{aho2016}, \cite{kim2018}, they are suitable for wind farm power tracking purposes on large farms. For instance, down-regulation providing power reserves were explored in \cite{fleming2016}.

%\emph{- Our novel contribution}

%Corrosion are uncontrollable in terms of the turbine's operation. However,

%\emph{- Structure of the paper}

% In this paper, we propose a closed-loop switching control architecture for individual damaged turbines that reduces thrust using down-regulation. The structural stresses are attenuated by regulating the estimated thrust force to a thrust reference of interest. The operator would have the flexibility on setting the appropriate thrust for each type of damage on the structure.
% The architecture, used on top of down-regulation controllers, presents expected thrust tracking performance with disturbance rejection and estimation error attenuation.  This would lead to a prevention on damage propagation in the structure, while power can still be generated. Consequently, the LCoE is reduced by generating power in faulty conditions, avoiding high maintenance costs, and extending the turbines' lifetime.   %derating as shown in \cite{richards2015}.

This paper proposes a switching control architecture for wind turbines that must satisfy a user-defined constraint on the fore-aft thrust forces. This architecture allows to track a demanded power profile when the thrust forces are lower than a given maximum allowed value. When such a value is reached, the proposed controller switches from tracking the demanded power to tracking the maximum allowed thrust force, thus satisfying the constraint on the turbine maximum loading.
In particular, the following contributions are presented:
\begin{enumerate}
    \item The extension of an existing power tracking controller via a switching thrust tracking feedback law that modifies the former's reference, by taking inspiration from the \emph{reference governor} literature \cite{garone2017reference};
    %\item the use of an online estimator for the thrust force;
    \item The introduction of an integral switching law for the thrust tracking controller in order to avoid chattering when the estimate thrust force lies on the constraint boundary;
    \item A large eddy simulation (LES) study showing the effectiveness of the proposed scheme for both laminar and turbulent wind speed profiles.
\end{enumerate}

The simulation study shows how the relative reduction in the thrust force results in a generated power reduction experienced by a given turbine. This allows to satisfy constraints on structural load, which can depend for instance on the current turbine state of structural health. Furthermore, the present paper will pave the way for the design of a wind farm power allocation control scheme, where healthy turbines can be assigned a larger demanded power to compensate for reductions in turbines that operate at their maximum allowed structural load.

The structure of this paper is as follows. First, the switching thrust tracking controller design is outlined in Section~\ref{thrust tracking}. Next, the down-regulation method is discussed in Section~\ref{drimplemented}.
Section~\ref{discus} presents simulation results of the proposed thrust tracking controller.
Finally, the paper is concluded in Section~\ref{conc}.

\section{Switching Control Architecture} \label{thrust tracking}
% This section presents a methodology to regulate the current average aerodynamic load for acting on individual turbines. The methodology is outlined by the feedback controller, the thrust computation and the controller design.

This section presents the proposed switching control architecture for load constraining. First an overview of the architecture will be given. Then, the individual constituting blocks, namely the thrust estimator, the thrust tracking controller, the switching law, and the demanded power down regulator will be described.

\subsection{Overview of the architecture}
The overall switching control architecture is presented in Figure~\ref{clthrust}, where we assume that external references $P^{\textrm{ref}}\in\mathbb{R}^+$ and $F_{T}^{\textrm{ref}}\in\mathbb{R}^+$ are provided for, respectively, the power to be generated by the given wind turbine and the maximum allowed fore-aft thrust forces. Starting from left, we can identify the following functional blocks

\begin{enumerate}
    \item The \emph{power down-regulator} $\mathcal{D}$, which can track a demanded power $P^{\textrm{dem}} \leq P^{\textrm{av}}$, where $P^{\textrm{av}}$ is the available aerodynamic power. $\mathcal{D}$ provides to the turbine a reference $\theta^{\textrm{ref}}$ and $\tau_\textrm{g}^{\textrm{ref}}$ for, respectively, the collective blade pitch and the generator torque (see Section~\ref{drimplemented}).
    \item The \emph{wind turbine}, which accepts as inputs the references $\theta^{\textrm{ref}}$ and $\tau_\textrm{g}^{\textrm{ref}}$ and provides the output measurements $\omega_{\textrm{r},\textrm{meas}}$, $\tau_\textrm{g, meas}$ and $\theta_{\textrm{meas}}$ representing the rotor speed, generator torque and the collective blade pitch angle, respectively.
    \item The \emph{wind speed estimator}, which uses the measurements $\omega_{\textrm{r},\textrm{meas}}$ and $\theta_{\textrm{meas}}$ to compute an estimate $\hat v_\textrm{w}$ of the rotor effective wind speed $v_\textrm{w}$. As in the present work we do not assume to have a reliable wind speed measurement, so the Immersion and Invariance (I\&I) wind speed estimator of \cite{liu2022} will be used.
    \item The \emph{thrust computation} block, which uses the wind speed estimate $\hat v_\textrm{w}$, $\omega_{\textrm{r},\textrm{meas}}$ and $\theta_{\textrm{meas}}$  to compute an estimate of the fore-aft thrust force $\hat F_\textrm{T}$.
    \item The \emph{mode switch} block, which together with the transfer function $K(s)$ constitutes the novel \emph{ switching controller} proposed in the present paper for tracking the thrust force. The output of $K(s)$ is the signal $u$, which is subtracted from the external reference $P^{\textrm{ref}}$ to obtain the demanded power $P^{\textrm{dem}} = P^{\textrm{ref}} - u$.
    
\end{enumerate}

In order to derive the proposed switching controller, we will introduce the following 
\begin{assumption}\label{as:1}
The down-regulation controller and wind speed estimator are asymptotically stable, as in \cite{aho2016, kim2018, liu2022}.
\end{assumption}
\begin{assumption}\label{as:2}  The dynamics from the demanded power to thrust force are represented by a first-order linear model as
\end{assumption}
\begin{equation} \label{thrustmodel}
    \frac{F_\mathrm{T}(s)}{P^\mathrm{dem}(s)}=\frac{A}{s+B}\overset{\operatorname{def}}{=}G(s).
\end{equation}

\begin{remark}
As Assumption \ref{as:1} holds, in steady conditions both generated power and thrust force will converge to a constant value. Short term transient behaviors in the turbine operation are not the focus of our controller. This justifies the use of Assumption \ref{as:2} for representing the dominant behavior.  
\end{remark}

We are now ready to introduce the control problem addressed by the present paper:

\begin{problem}\label{prob:mainProblem}
Design a feedback control law that extends, rather than replacing, the existing demanded power controller $\mathcal D$ such that:
\begin{enumerate}
    \item the fore-aft thrust force $F_\textrm{T}$ does not exceed a user defined upper bound $\bar F_\textrm{T}$;
    \item the tracking performance of the desired power reference $P^{\textrm{ref}}$ is unaltered whenever $F_\textrm{T} \leq \bar F_\textrm{T}$.
\end{enumerate}
\end{problem}

The next subsection introduces the novel \emph{switching controller} developed in this paper.

\subsection{Switching Controller Design}
% The controller is proposed based on a set of assumptions that are consistent with the high-fidelity simulation results.

This section describes the design of the Mode Switch and the transfer function $K(s)$ introduced in previous subsection, towards solving the control problem posed in Problem~\ref{prob:mainProblem}.

The first observation is that, in order to satisfy point~2 in Problem~\ref{prob:mainProblem}, it is sufficient that the signal $u$ becomes zero, and thus $P^{\textrm{dem}} = P^{\textrm{ref}}$. In order to do this, the Mode Switch will be designed such that the lower feedback loop in Figure~\ref{clthrust} will be open whenever the constraint on the thrust force is not exceeded. In particular, the switching will be defined by introducing the following signal 
\begin{equation}
\tilde e = \begin{cases} e, &\text{if } e < 0 \ \text{ or } e_\mathrm{I} < 0 \\
        0, & \text{otherwise}
        \end{cases}
\end{equation}
where $e = F_\textrm{T}^{\mathrm{ref}} - \hat{F}_\mathrm{T}$ and $e_\mathrm{I} = \int_0^t{e(\tau)} \textrm{d}\tau$ represent, the difference between the estimated thrust force and its reference, herein set as equal to the upper bound, i.e.  $F_\textrm{T}^{\mathrm{ref}}$  = $\bar F_\textrm{T}$, and the time integral of that, respectively. The rationale for this definition with the inclusion of the integral term is to avoid \emph{chattering} when the thrust force is close to its reference, as is done for instance in the literature on Integral Sliding Mode control \cite{laghrouche2007higher}.

When the mode switch is active, that is when $\tilde e \neq 0$, the lower feedback loop involving the transfer function $K(s)$ is closed. The design of $K(s)$, outlined in the subsequent part, will thus determine the fulfillment of requirement~1 in Problem~\ref{prob:mainProblem}.

The approach is to design  $K(s)$ such that $\hat{F}_\mathrm{T}$ will track its reference $F_\textrm{T}^{\mathrm{ref}}$. From this point of view, notice that variations of the power reference and wind speed also contribute to the thrust response. As such, $P^\mathrm{ref}$ and $v_\mathrm{w}$ are considered disturbances acting into the system. This said, the thrust response can be written in the Laplace domain as
\begin{equation} \label{plantoutput}
 F_\mathrm{T} (s) = - G(s)u(s) + G_\mathrm{d, \, 1}(s) P^\mathrm{ref}(s) + G_\mathrm{d, \, 2}(s) v_\mathrm{w} (s),
\end{equation}
\begin{flalign} \label{controlinput}
\mathrm{where } \, \, \, \, \, \ u(s)  = - K(s) \tilde e(s) = - K(s) \left( F_\mathrm{T}^\mathrm{ref}(s) -F_\mathrm{T} (s) - n(s) \right). &&
\end{flalign}
%As being $P^\mathrm{ref}$ considered a disturbance, $G_\mathrm{d, \, 1}(s) = G(s)$.  
$G(s)$ represents the transfer function of the join \emph{power down-regulator} and \emph{wind turbine} from the considered demanded power $P^{\mathrm{dem}}$ input and the thrust $F_\mathrm{T}$ output. $G_\mathrm{d, \, 1}(s) = G(s)$ from $P^\mathrm{ref}$ to $F_\mathrm{T}$ and $G_\mathrm{d, \, 2}(s)$ from $v_\mathrm{w}$ to $F_\mathrm{T}$ are the transfer functions from the considered disturbances on the down-regulated turbine. Furthermore, $n(s)= F_\mathrm{T}(s) - \hat{F}_\mathrm{T}(s)$ represents the estimation error between the true thrust and the estimated one (cfr. Section~\ref{sub:thrust}). 

Substituting Eq. \eqref{controlinput} into Eq. \eqref{plantoutput}, and reorganizing the terms we have the closed-loop response as
\begin{equation}  \label{closeloopoutput}
\begin{array}{cc}
 F_\mathrm{T} (s) & = \dfrac{K(s)G(s)}{1+K(s)G(s)} F_\mathrm{T}^\mathrm{ref}(s)  + \dfrac{G_\mathrm{d, \, 1}(s)}{1+K(s)G(s)} P^\mathrm{ref}(s) \\ & 
 \\
     & +\dfrac{G_\mathrm{d, \, 2}(s)}{1+K(s)G(s)}  v_\mathrm{w} (s)  - \dfrac{K(s)G(s)}{1+K(s)G(s)} n(s).
\end{array}
\end{equation}

The controller $K(s)$ should be designed to stabilize the poles of the closed-loop system, i.e. the roots of $1+K(s)G(s)=0$, and to guarantee steady-state convergence from changes in the reference, disturbances and estimation error.  A PI controller is therefore chosen in this paper as an effective way to track the reference, to reject the disturbances and to attenuate the estimation error
\begin{equation}\label{picontrol}
    K(s) = K_\mathrm{P} + \frac{K_\mathrm{I}}{s}.
\end{equation}

The closed-loop characteristic equation is then equal to
\begin{equation}
    s^2 + (B + K_\mathrm{P} A)s + K_\mathrm{I} A =  s^2 + 2 \zeta \omega_\mathrm{n} s + \omega_\mathrm{n}^2 = 0.
\end{equation}
%= T(s) F_T^{ref}(s) - S_1(s) P^{ref}(s)  - S_2(s) v_w (s) - T(s) n(s)

\begin{proposition}
The steady-state thrust tracking error is bounded for steps in reference, disturbances and estimation error.
\end{proposition}

Using Eq. \ref{closeloopoutput}, we can derive the thrust tracking error function
\begin{equation}
\begin{array}{cl}
    E(s) & = F_\mathrm{T}(s)-F_\mathrm{T}^\mathrm{ref}(s) \\
     & = \underbrace{\left(T(s)-1 \right) F_\mathrm{T}^\mathrm{ref}(s)}_{E_1(s)} + \underbrace{S(s) G_\mathrm{d, \,1} (s) P^\mathrm{ref} (s)}_{E_2(s)} \\
     & \, \, \, \, + \underbrace{S(s) G_\mathrm{d, \,2} (s) v_\mathrm{w} (s)}_{E_3(s)} - \underbrace{T(s) n(s)}_{E_4(s)} \,, 
\end{array}
\end{equation} 
where $S(s)=1/(1+K(s)G(s))$ and $T(s)= K(s)G(s)/(1+K(s)G(s))$ are the sensitivity and complementary sensitivity functions, respectively.

By the final value theorem, the state steady errors due to steps on $F_T^\mathrm{ref}$, $P^\mathrm{ref}$, $v_\mathrm{w}$, and $n$ can be computed for the proposed controller as

\begin{equation}
\begin{array}{cl}
   e_{\mathrm{ss},i} & = \lim_{t \to \infty} e_i(t)  \\
   & = \lim_{s \to 0} s E_i(s),
\end{array}
\end{equation}
where $i=1,2,3,4$.

%Note that, from Eq. \eqref{thrustmodel} and \eqref{picontrol}, $T(0)=1$, and $S(0)=0$. Also $G_\mathrm{d, \, 1}(s)$ and $G_\mathrm{d, \, 2}(s)$ are considered to be also type 0, such that $\lim_{s \to 0}  S(s) G_\mathrm{d, \, 1}(s) =\lim_{s \to 0}  S(s) G_\mathrm{d, \, 2}(s)=0$.
\begin{comment}
\begin{equation}
\begin{array}{cl}
   e_\mathrm{ss_1} & = \lim_{s \to 0} s E_1(s) \\
    & = \lim_{s \to 0} s \left( T(s) - 1 \right) \frac{1}{s}  \\
    & = T(0) - 1 \\
    & = 0,
\end{array}
\end{equation}
\begin{equation}
\begin{array}{cl}
   e_\mathrm{ss_2} & = \lim_{s \to 0} s E_2(s) \\
    & = \lim_{s \to 0} s S(s) G_\mathrm{d, \, 1} (s) \frac{1}{s}  \\
    & = \lim_{s \to 0} S(s) G_\mathrm{d, \, 1} (s) \\
    & = 0,
\end{array}
\end{equation}
\begin{equation}
\begin{array}{cl}
   e_\mathrm{ss_3} & = \lim_{s \to 0} s E_3(s) \\
    & = \lim_{s \to 0} s S(s) G_\mathrm{d, \, 2} (s) \frac{1}{s}  \\
     & = \lim_{s \to 0} S(s) G_\mathrm{d, \, 2} (s) \\
    & = 0,
\end{array}
\end{equation}
\begin{equation}
\begin{array}{cl}
   e_\mathrm{ss_4} & = \lim_{s \to 0} s E_4(s) \\
    & = \lim_{s \to 0} s T(s) \frac{1}{s}  \\
    & = T(0) \\
    & = 1.
\end{array}
\end{equation}
\end{comment}
By Assumption~\ref{as:1}, $G_\mathrm{d, \, 1}(s)$ and $G_\mathrm{d,\,2}(s)$ are stable functions, as well as the estimation error $n$ is bounded. Therefore, the steady state error vector is $e_{\mathrm{ss}}=[0, \, 0, \, 0, \, -T(0)]$ for steps on $F_T^\mathrm{ref}$, $P^\mathrm{ref}$, $v_\mathrm{w}$, and $n$.

\begin{remark}
The controller can be calibrated with an offset in to the thrust reference to take into account $e_\mathrm{ss, 4}$ if a constant bias in the thrust estimation exists. In addition, extra measurement devices can be used to improve the estimations and reduce the bias, such as the use of strain-gauges or accelerometers in the tower structure.
\end{remark}
 
\begin{remark}
Deviations in the model from Assumption \ref{as:2} can occur at different wind speeds given the non-linearity of the turbines, therefore it might degrade the designed controller performance. %As for instance, gain-scheduled techniques can be further explored, not done herein for the sake of simplicity. 
Nevertheless, as Assumption \ref{as:1} and monotonicity from $\mathcal{D}$ still hold, the tracking is kept, as well as steps from disturbances rejected and from noise attenuated, even though model uncertainties exist. 
\end{remark}

%By the final value theorem,

\subsection{Thrust Computation} \label{sub:thrust}

The estimated average fore-aft thrust force, representing the aerodynamic loads, is computed by
\begin{equation}\label{thrusteq}
\hat{F}_\mathrm{T}=0.5\rho \pi R^2 {\hat{v}_\mathrm{w}^2} C_\mathrm{T} \left( \frac{R\omega_{\textrm{r},\textrm{meas}}}{\hat{v}_\mathrm{w}}, \theta_{\textrm{meas}} \right) ,
\end{equation}
where $\hat{v}_\mathrm{w}$ is the estimated effective wind speed, $\rho$ the air density, $R$ the rotor radius and $C_\mathrm{T}$ the thrust coefficient computed with the measured rotor speed $\omega_{\textrm{r},\textrm{meas}}$ and collective blade pitch angle $\theta_{\textrm{meas}}$. In this work, the estimation of the effective wind speed is obtained from the I\&I estimator \cite{liu2022} through measurements of rotor speed, generator torque and blade pitch angles. 

%The relation between the thrust force and the bending moment on the tower base can be roughly considered for calculate the higher structure stresses from aerodynamic loads. As larger structures in length are being used, larger are the associated stresses.

%Since it is computed as the average loads acting on the blades of the turbine, 

%Also, high accuracy is not required, since it represents the average loads acting on the blades of the turbine.

\subsection{Power down-regulator} 
 \label{drimplemented}

\begin{figure}[t!]
\centering
\includegraphics[width=\linewidth]{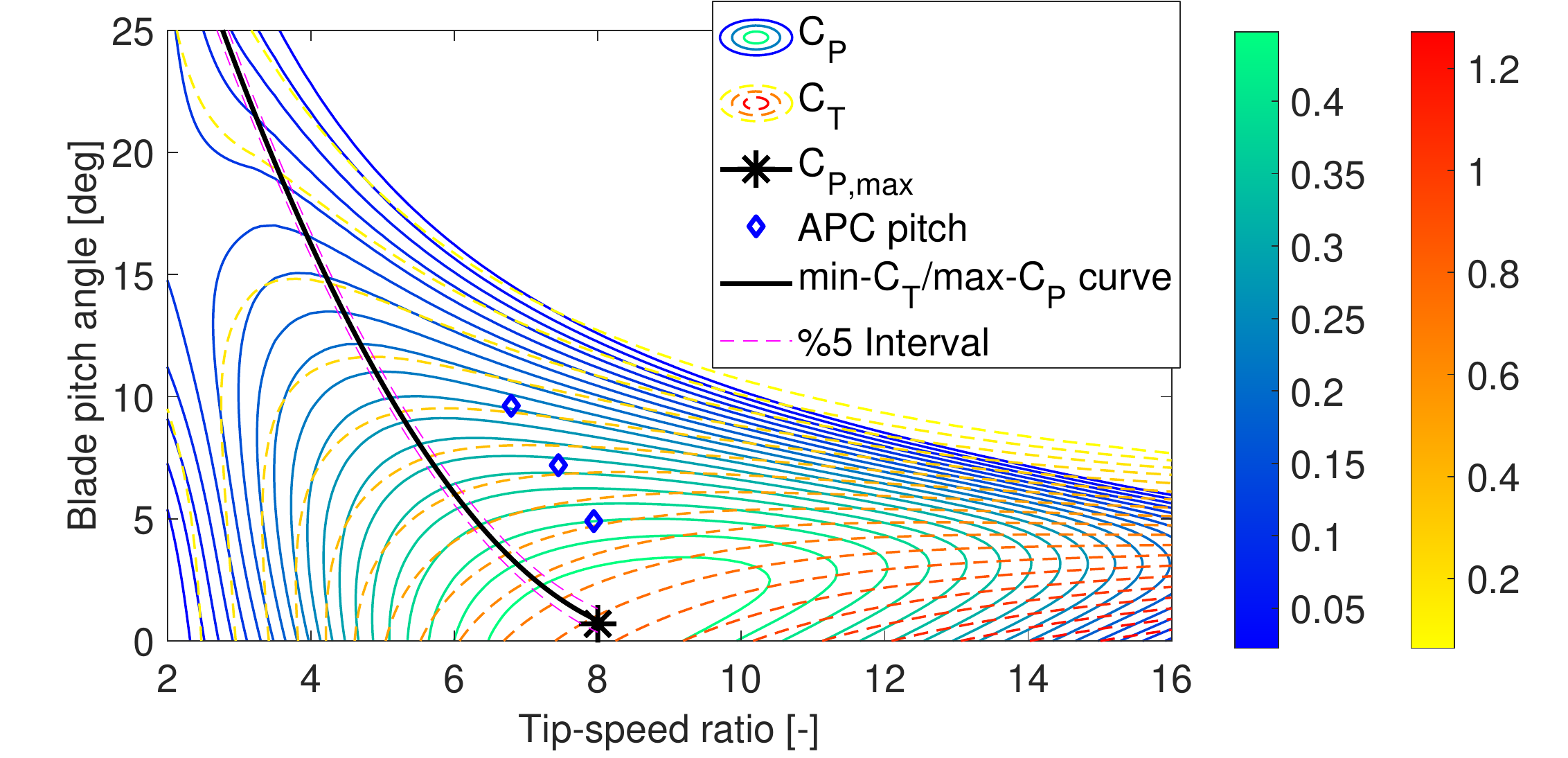}
\caption{C$_\mathrm{P}$ and C$_\mathrm{T}$ of the operation points of ``APC pitch" given a reduction in the demanded power (5, 4 and 3 MW) at 9 m/s wind speed  and min-C$_\mathrm{T}$/max-C$_\mathrm{P}$ curve for the DTU 10MW reference wind turbine.}
\label{Cp and Ct}
\end{figure}

For the down-regulator $\mathcal D$ the APC pitch approach~\cite{aho2016,kim2018} is adopted in this work, because of the following characteristics:
\begin{itemize}
    \item A monotonic thrust reduction in response to monotonic demanded power reduction is achieved;
    \item Operation close to the min-$C_T$ method, see Fig.~\ref{Cp and Ct};
    \item Stability margin for the stall region \cite{deshpande2012}.
\end{itemize}

The turbine is down-regulated using the blade pitch controller, which consists of a gain-scheduled PI control law
\begin{equation}\label{pitchlaw}
\begin{array}{cc}
    \theta = & \overline{K}_\mathrm{P}(\theta_\mathrm{meas})\left[ \omega_\mathrm{g, \, meas} - \omega_\mathrm{g}^{\mathrm{ref}}(P^\mathrm{dem}) \right] \\
     & \, \, \, \, + \dfrac{\overline{K}_\mathrm{I} (\theta_\mathrm{meas})}{s} \left[ \omega_\mathrm{g, \, meas} - \omega_\mathrm{g }^{\mathrm{ref}}(P^\mathrm{dem}) \right],
\end{array}
\end{equation}
where $\omega_\mathrm{g, \, meas}$ and $\theta_\mathrm{meas}$ are the measured generator speed and the measured collective blade pitch angle, respectively.  The reference generator speed as function of the demanded power $P^\mathrm{dem}$ is represented by $\omega_\mathrm{g}^{\mathrm{ref}}(P^\mathrm{dem})$. $\overline{K}_\mathrm{P}(\theta_\mathrm{meas})$ and $\overline{K}_\mathrm{I}(\theta_\mathrm{meas})$ are the gain-scheduled proportional and integral gains \cite{ROSCO2021}. 
%Also, a low-pass filter with bandwidth of 0.4 rad/s  is included on the blade pitch measure used on the gain-scheduled pitch controller to avoid chattering due to the large sampling time (0.1 s).
\begin{comment}
\begin{figure}[h!]
\centering
\includegraphics[width=0.8\linewidth]{Figures/Lookup_Table_P_omega_DTU10MW_ROSCO_constant_power.pdf}
\caption{Generator speed reference as function of the demanded power for the DTU 10MW RWT}
\label{powerdemandedcurve}
\end{figure}
\end{comment}
While the generator torque controller is applied to track the demanded power by multiplying it by the inverse of the measured generator speed as
\begin{equation}\label{torquelaw}
    \tau_\mathrm{g} = \frac{P^\mathrm{dem}} {\eta_\mathrm{eff} \omega_\mathrm{g, \, meas}}.
\end{equation}

Both controllers are applied whenever the demanded power is lower than the rated power, and the measured generator speed is higher than the reference generator speed or the blade pitch angle is higher than a switch blade pitch angle. Otherwise, the turbine follows the conventional turbine controllers \cite{ROSCO2021}.

\section{SIMULATIONS AND DISCUSSIONS} \label{discus}

The proposed control architecture is evaluated in the high-fidelity simulation environment, Simulator for Wind Farm Applications (SOWFA), developed by the National Renewable Energy Laboratory \cite{churchfield2012}. %The controllers are implemented in MATLAB through a network-based communication interface \cite{doekemeijer2019}. 
SOWFA implements the actuator line method, embedded in a computational fluid dynamics flow, accounting for the Coriolis force and Buoyancy effects. Capturing turbine interactions, SOWFA was chosen aiming for future wind farm investigations.  The DTU 10MW reference wind turbine is used in this work \cite{bak2013}. The control parameters of the controllers are based on the values provided with the NREL's Reference OpenSource Controller (ROSCO) \cite{ROSCO2021}. An overview of the parameters for the SOWFA simulations in this paper is shown in Table \ref{sowfapar}.

\begin{table}[h]
\caption{SOWFA simulation parameters}
\label{sowfapar}
\begin{center}
\begin{tabular}{l c}
\hline
Property  & Value \\
\hline
Sub-grid-scale model & One-equation eddy viscosity\\
Domain size & 3 km $\times$ 3 km $\times$ 1 km\\
Cell size outer regions & 10 m $\times$ 10 m $\times$ 10 m\\
Cell size near rotor & 2.5 m $\times$ 2.5 m $\times$ 2.5 m\\
 Simulation timestep & 0.04 s \\
 Atmospheric boundary layer stability & Neutral \\
 Mean inflow wind speed & 9 m/s\\
 Turbulence intensity & 0.0 and 5.0 \% \\
 Turbine rotor approximation & Actuator Line Model \\
 Turbine type & DTU 10 MW \\
 Turbine rotor diameter & 178.3 m \\
 Turbine hub height & 119 m \\
 Blade smearing factor & 5.0 m  \\
\hline
\end{tabular}
\end{center}
\end{table}

\begin{figure}[b!]
\centering
\includegraphics[width=\linewidth]{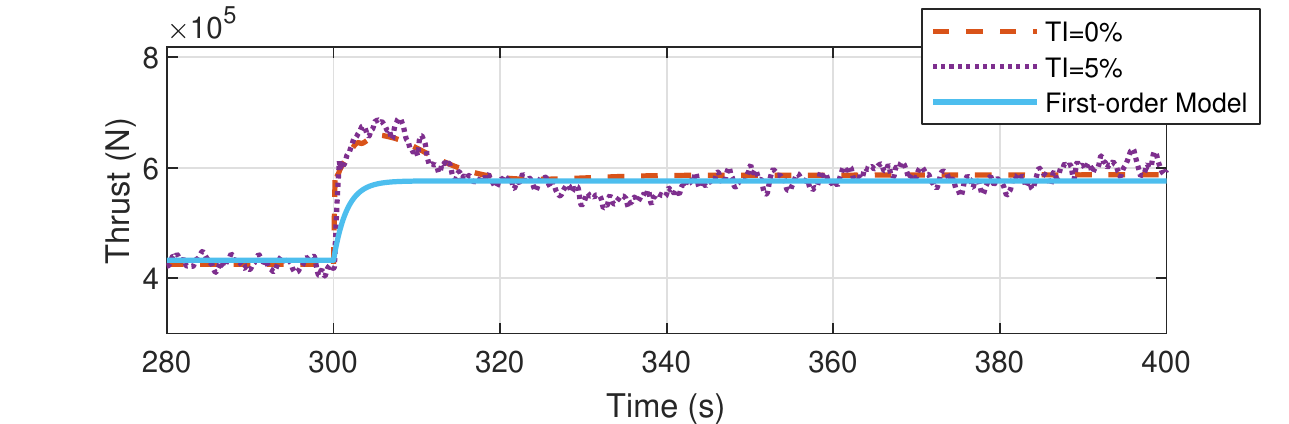}
\caption{Thrust response of a step demanded power from 3 to 4 MW at 9 m/s with the down-regulator  $\mathcal D$ defined in Section \ref{drimplemented}.}
\label{stepresponse}
\end{figure}

\begin{figure}[h!]
\centering
\includegraphics[width=0.7\linewidth]{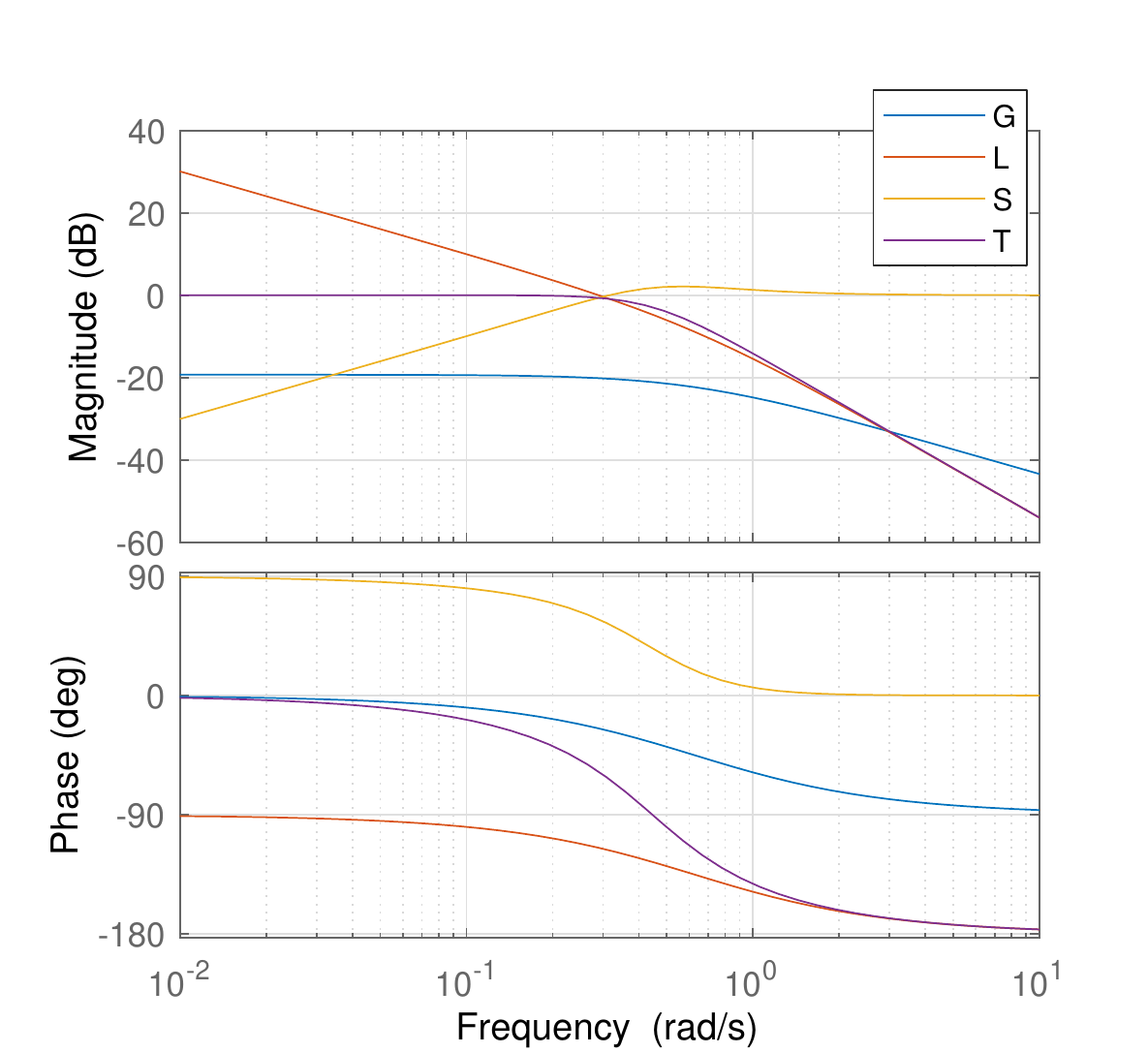}
\caption{Bode magnitude and phase plots of L(s)=G(s)K(s), S(s) and T(s) for G(s) = 0.067625/s+0.625 and K(s)=2.947/s. The crossover frequencies 0.2 Hz and 7.16 Hz of L at +20dB and -20dB are considerable high and low, respectively. Meaning high tracking robustness and disturbance rejection \cite{skogestad2005}.}
\label{bode}
\end{figure}

At a mean wind speed of 9 m/s (i.e. below rated wind speed), the first-order model from Eq. \eqref{thrustmodel} was identified from down-regulating 3 to 4 MW of power using the thrust step response, see Fig. \ref{stepresponse}. The model parameters were identified as $A=0.068$ and $B=0.625$. By considering a simple first-order model, only the dominant dynamic behavior is captured.
For the controller design, we choose $K_\mathrm{P}=0$ and a damping coefficient $\zeta=0.7$, then $\omega_\mathrm{n}=0.446$ and $K_\mathrm{I}=2.947$. As is seen in Fig. \ref{bode}, the integrator is sufficient to add tracking and robustness to the plant from $G$ to $L=KG$, where the disturbances are reduced
and the estimation error is attenuated, as shown by $S$ and $T$. The controller provides an infinity gain margin and a phase margin of 65.2 degrees at frequency 0.047 Hz for the identified model. Up to 99\% of tracking until frequencies of 0.2 Hz and at least 90.9\% of reduction for frequencies greater than 7.16 Hz (from the crossover frequencies of L at $\pm$ 20 dB). %Moreover, the integrator is sufficient to guarantee the steady state error of $F_\mathrm{T} - F_\mathrm{T}^\mathrm{ref}$ to be bounded given step changes on the disturbances and estimation error.
 
\begin{figure}[b!]
\centering
\includegraphics[width=\linewidth]{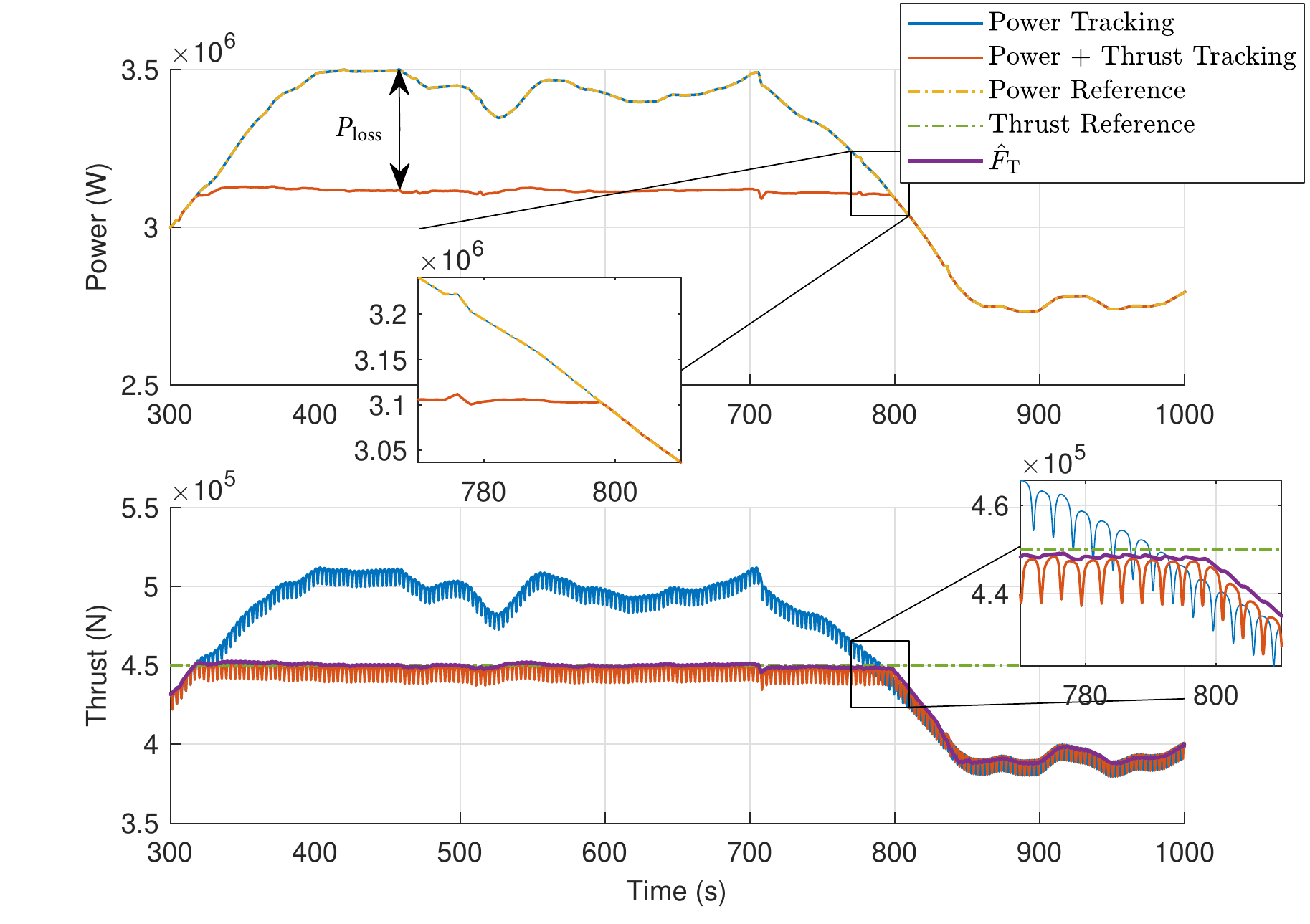}
\caption{Power and thrust of the simulations with no turbulence. Thrust tracking with reference of 450 kN.}
\label{resultsfig}
\end{figure}

\begin{figure}[h!]
\centering
\includegraphics[width=\linewidth]{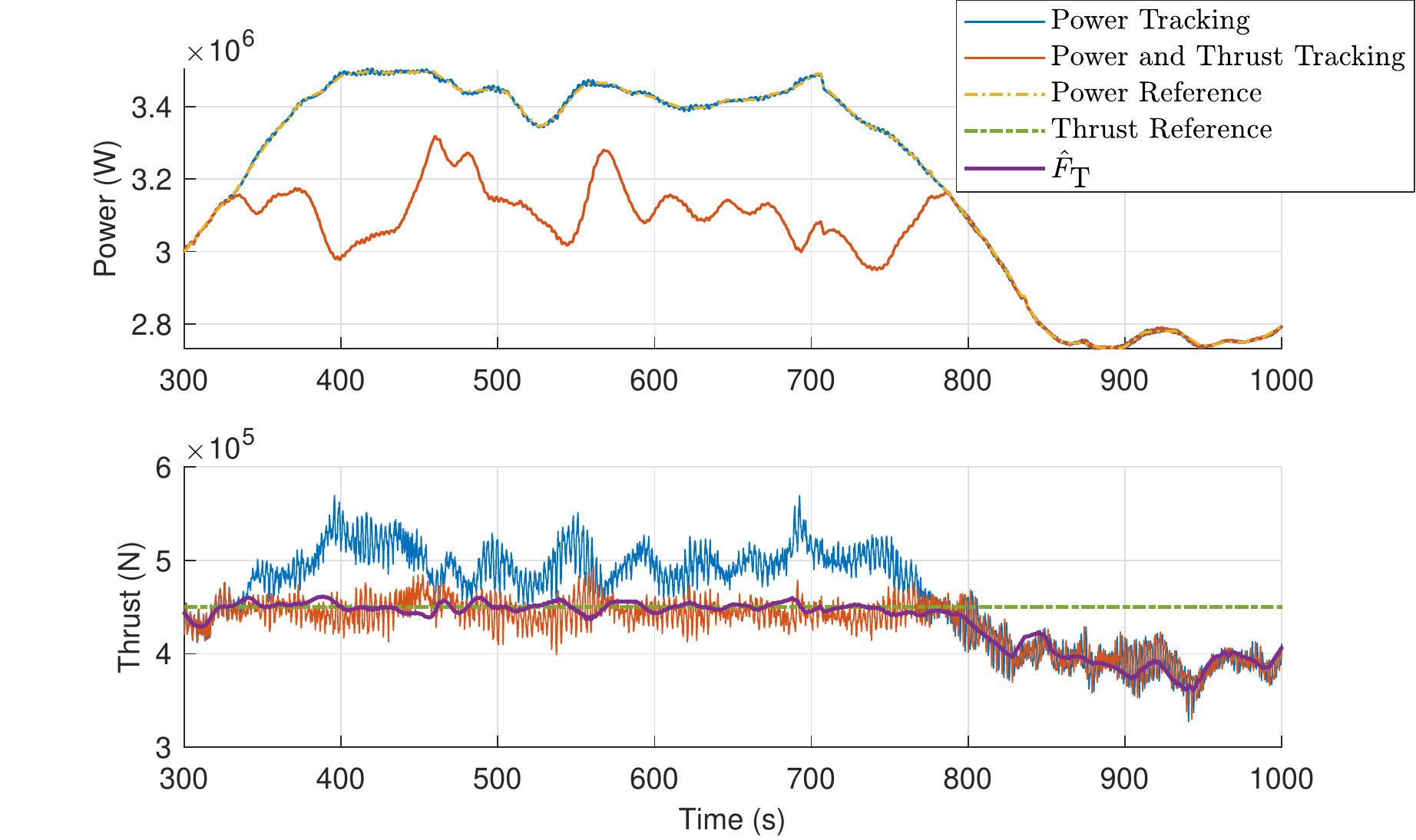}
\caption{Power and thrust of the simulations with turbulence (turbulence intensity of 5\%). Thrust tracking with reference of 450 kN}
\label{resultsfig2}
\end{figure}

The simulations were performed by feeding a reference power signal below the available power based on the normalized standard test signal \cite{pilong2013}. The results of the thrust tracking controller are presented in Figs. \ref{resultsfig} and \ref{resultsfig2}, using an uniform wind inflow and a turbulent one, respectively.
As the maximum thrust is about 512 kN by just tracking power, the thrust references were accordingly set as 500, 475, 450, 425 kN for the uniform inflow case.
The results shown in Table \ref{tablepowerloss} present a rough linear relation between the percentage of the maximum power loss and the percentage of reduction of thrust in the case studies. %In the turbulent scenario, because of thrust and oscillations increase, we adopted the thrust reference of 500 kN to represent the effectiveness of the framework for the turbulent wind case.

%In order to reduce oscillations on the power generation of the down-regulation in the turbulent wind inflow simulation, the bandwidth of the applied low-pass filter on the rotor speed measures at sampling time of 0.1 s is reduced from 1.1297 rad/s \cite{ROSCO2021} to 0.1 rad/s. The drawback of excessive filtering is a delay on tracking the demanded power. Also, a low-pass filter with bandwidth of 0.4 rad/s  is included on the blade pitch measure used on the gain-scheduled pitch controller to avoid chattering. %Note that filter designs depends on the sampling time of the signal.

\begin{table}[h]
\caption{Power loss as function of the thrust reference}
\label{tablepowerloss}
\begin{center}
\begin{tabular}{c c c c}
\hline
Thrust  & Percentage of & Maximum   & Percentage of \\
reference & thrust reduction & power loss & power reduction \\ $\mathrm{[kN]}$ & $[\%]$ & $\mathrm{[MW]}$ &  $[\%]$ \\
\hline
500 & 2.35 & 0.0750 & 2.1\\
475 & 7.23 & 0.2175 & 6.2\\
450 & 12.12 & 0.3826 & 10.9\\
425 & 16.99 & 0.5382 & 15.4 \\
\hline
\end{tabular}
\end{center}
\end{table}

The proposed control architecture was able to track the computed thrust sufficiently, even though oscillations appear in the thrust force. These oscillations are due to the tower effect characterized by the blade passing frequency (3P) and to the inflow wind profile reproduced by SOWFA. They are partially accommodated by the adopted wind speed estimator and thrust computation. In addition, the proposed integral switching law presents robustness against chattering and smooth transitions. %Although the stability margin from stall regions of the power down-regulator might not be enough in extreme turbulent scenarios, the proposed control architecture presents the expected good performance on thrust limitation, independently of the possible issue of the down-regulator.

\section{CONCLUSIONS} \label{conc}

A closed-loop switching control architecture for addressing aerodynamic load constraints on wind turbines is presented. The framework makes use of down regulation and wind speed estimations. The proposed architecture effectively limits the aerodynamic loads, where the appropriate user-defined constraint would ensure safety on the system. In fact, such approaches present an undesirable loss in power production but it can be acceptable in a wind farm wide perspective. Therefore, the proposed architecture aims to prevent mechanical structures from failures and their associated costs by avoiding high stresses from the aerodynamic loads. Also, this can lead to a profit from keeping turbines operating in non-designed conditions, such as corroded structures. 

%This would allow the further operation of offshore turbines with degraded structures, where structural damage propagation is then decreased.  

Future work will focus on the development and the performance of model predictive controllers for down-regulation purposes, aiming to avoid unstable flow behaviors and oscillations in response signals, and expand the present framework to farm simulations. 

                            
%%%%%%%%%%%%%%%%%%%%%%%%%%%%%%%%%%%%%%%%%%%%%%%%%%%%%%%%%%%%%%%%%%%%%%%%%%%%%%%%
%\section*{APPENDIX}

\section*{ACKNOWLEDGMENT}

The authors would like to acknowledge the WATEREYE project (grant no. 851207). This project has received funding from the European Union Horizon 2020 research and innovation programme under the call H2020-LC-SC3-2019-RES-TwoStages.
%The preferred spelling of the word  acknowledgment  in America is without an  e  after the  g . Avoid the stilted expression,  One of us (R. B. G.) thanks . . .   Instead, try  R. B. G. thanks . Put sponsor acknowledgments in the unnumbered footnote on the first page.

%%%%%%%%%%%%%%%%%%%%%%%%%%%%%%%%%%%%%%%%%%%%%%%%%%%%%%%%%%%%%%%%%%%%%%%%%%%%%%%%
\bibliographystyle{IEEEtran}
\bibliography{IEEEabrv,IEEEexample}

\end{document}